\documentclass[12pt]{article}
\usepackage{amsmath,color}
\usepackage[dvips]{graphicx}
\input amssym
\input amssym.def

\def\red#1{{\color{red} #1}}
\begin{document}

%%%%%%%%%%%%%%%%%%%%%%%%%%%%%%%%%%%%%%%%%%%%%%%%%%%%%%%%%%%%%%%%%%%%%
\def\prg#1{\medskip\noindent{\bf #1}}  \def\ra{\rightarrow}
\def\lra{\leftrightarrow}              \def\Ra{\Rightarrow}
\def\nin{\noindent}                    \def\pd{\partial}
\def\dis{\displaystyle}
\def\grl{{GR$_\Lambda$}}               \def\Lra{{\Leftrightarrow}}
\def\cs{{\scriptstyle\rm CS}}          \def\ads3{{\rm AdS$_3$}}
\def\Leff{\hbox{$\mit\L_{\hspace{.6pt}\rm eff}\,$}}
\def\bull{\raise.25ex\hbox{\vrule height.8ex width.8ex}}
\def\ric{{Ric}}                        \def\tric{{(\widetilde{Ric})}}
\def\tmgl{\hbox{TMG$_\Lambda$}}        \def\phb{\phantom{\Big|}}
\def\Lie{{\cal L}\hspace{-.7em}\raise.25ex\hbox{--}\hspace{.2em}}
\def\sS{\hspace{2pt}S\hspace{-0.83em}\diagup}   \def\hd{{^\star}}
\def\dis{\displaystyle}                \def\ul#1{\underline{#1}}
\def\mb#1{\hbox{{\boldmath $#1$}}}     \def\grp{{GR$_\parallel$}}
\def\irr#1{^{(#1)}}                    \def\ric{{Ric}}

%% \hook is a better version of \rfloor
\def\hook{\hbox{\vrule height0pt width4pt depth0.3pt
\vrule height7pt width0.3pt depth0.3pt
\vrule height0pt width2pt depth0pt}\hspace{0.8pt}}
\def\semidirect{\;{\rlap{$\supset$}\times}\;}
\def\bm#1{\hbox{{\boldmath $#1$}}}
\def\ir#1{{}^{(#1)}}  \def\inn{\hook}

\def\G{\Gamma}         \def\S{\Sigma}        \def\L{{\mit\Lambda}}
\def\D{\Delta}         \def\Th{\Theta}
\def\a{\alpha}         \def\b{\beta}         \def\g{\gamma}
\def\d{\delta}         \def\m{\mu}           \def\n{\nu}
\def\th{\theta}        \def\k{\kappa}        \def\l{\lambda}
\def\vphi{\varphi}     \def\ve{\varepsilon}  \def\p{\pi}
\def\r{\rho}           \def\Om{\Omega}       \def\om{\omega}
\def\s{\sigma}         \def\t{\tau}          \def\eps{\epsilon}
\def\nab{\nabla}       \def\btz{{\rm BTZ}}   \def\heps{\hat\eps}
\def\bt{{\bar t}}      \def\br{{\bar r}}     \def\bth{{\bar\theta}}
\def\bvphi{{\bar\vphi}}  \def\bO{{\bar O}}   \def\bx{{\bar x}}
\def\by{{\bar y}}      \def\bom{{\bar\om}}
\def\tphi{{\tilde\vphi}} \def\tt{{\tilde t}}

\def\tG{{\tilde G}}   \def\cF{{\cal F}}      \def\bH{{\bar H}}
\def\cL{{\cal L}}     \def\cM{{\cal M }}     \def\cE{{\cal E}}
\def\cH{{\cal H}}     \def\hcH{\hat{\cH}}
\def\cK{{\cal K}}     \def\hcK{\hat{\cK}}    \def\cT{{\cal T}}
\def\cO{{\cal O}}     \def\hcO{\hat{\cal O}} \def\cV{{\cal V}}
\def\cE{{\cal E}}     \def\cR{{\cal R}}      \def\hR{{\hat R}{}}     \def\hL{{\hat\L}}     \def\tom{{\tilde\omega}}
\def\tb{{\tilde b}}   \def\tA{{\tilde A}}    \def\tv{{\tilde v}}
\def\tT{{\tilde T}}   \def\tR{{\tilde R}}    \def\tcL{{\tilde\cL}}
\def\hy{{\hat y}\hspace{1pt}}  \def\tcO{{\tilde\cO}}
\def\bA{{\bar A}}     \def\bB{{\bar B}}      \def\bC{{\bar C}}
\def\bG{{\bar G}}     \def\bD{{\bar D}}      \def\bH{{\bar H}}
\def\bK{{\bar K}}     \def\bL{{\bar L}}

\def\rdc#1{\hfill\hbox{{\small\texttt{reduce: #1}}}}
\def\chm{\checkmark}  \def\chmr{\red{\chm}}
%%%%%%%
\def\nn{\nonumber}                    \def\vsm{\vspace{-9pt}}
\def\be{\begin{equation}}             \def\ee{\end{equation}}
\def\ba#1{\begin{array}{#1}}          \def\ea{\end{array}}
\def\bea{\begin{eqnarray} }           \def\eea{\end{eqnarray} }
\def\beann{\begin{eqnarray*} }        \def\eeann{\end{eqnarray*} }
\def\beal{\begin{eqalign}}            \def\eeal{\end{eqalign}}
\def\lab#1{\label{eq:#1}}             \def\eq#1{(\ref{eq:#1})}
\def\bsubeq{\begin{subequations}}     \def\esubeq{\end{subequations}}
\def\bitem{\begin{itemize}}           \def\eitem{\end{itemize}}
\renewcommand{\theequation}{\thesection.\arabic{equation}}

%%%%%%%%%%%%%%%%%%%%%%%%%%%%%%%%%%%%%%%%%%%%%%%%%%%%%%%%%%%%%%%%%%%%%%%%%%%%%%%
\title{Entropy in general relativity: Kerr-AdS black hole}

\author{M. Blagojevi\'c and B. Cvetkovi\'c\footnote{
        Email addresses: {mb@ipb.ac.rs, cbranislav@ipb.ac.rs}} \\
                          Institute of Physics, University of Belgrade,\\
                      Pregrevica 118, 11080 Belgrade-Zemun, Serbia}
%\date{\today}
\maketitle

\begin{abstract}
A general Hamiltonian approach to black hole thermodynamics is used to study entropy and conserved charges for Kerr-AdS solutions in general relativity. These thermodynamic variables are first consistently defined by choosing suitable boundary conditions, and then, they are shown to satisfy the first law of black hole dynamics
\end{abstract}

%%%%%%%%%%%%%%%%%%%%%%%%%%%%%%%%%%%%%%%%%%%%%%%%%%%%%%%%%%%%%%%%%%%%%%%%%%%%%%%
\section{Introduction}
\setcounter{equation}{0}

Conserved charges of a gravitating system as a whole, energy-momentum and angular momentum, are deeply connected to the asymptotic behaviour of the gravitational field. Early investigations of these concepts in general relativity were focussed on asymptotically flat spacetimes, see for instance \cite{x1,x2}. In the mid of 1980s, the conserved charges were defined and calculated also for asymptotically anti-de Sitter (AdS) solutions \cite{x3}. Although these quantities are the subject in their own right, they also play an essential role in black hole thermodynamics. In the early 1990s, an important progress has been made by interpreting black hole entropy, in \emph{asymptotically flat} spacetimes, as the gravitational charge on horizon \cite{x4,x5}. However, the relation between entropy, conserved charges and angular velocity for \emph{asymptotically AdS} black holes, studied more intensively from the end of 1990s, turns out to be a subject with controversial aspects, see Gibbons \cite{x6}.

Recently, a Hamiltonian approach to entropy has been proposed \cite{x7}, describing conserved charges and entropy for any black hole as canonical charges at infinity and on horizon, respectively. It was applied to Schwarzschild-AdS black holes with or without torsion \cite{x7}, asymptotically flat Kerr black holes \cite{x8} and to three-dimensional Kerr-AdS black holes \cite{x9}. The approach worked so well that we were tempted to go on and analyze a more complicated case of the four-dimensional Kerr-AdS black hole. In the present paper, we use it to study thermodynamic variables and the first law for Kerr-AdS solution in general relativity with a cosmological constant (\grl).

The paper is organized as follows. In section \ref{sec2}, we review basic aspect of the Hamiltonian approach to entropy. In section. \ref{sec3}, we describe geometry of the Kerr-AdS black hole in the tetrad formalism. In section \ref{sec4}, we use the approach proposed in \cite{x7} to calculate energy and angular momentum, and then, in section \ref{sec5}, to derive the expression for entropy and verify the first law of black hole dynamics. In section \ref{sec6}, we discuss some controversial issues regarding the first law, and Appendix \ref{appA} is devoted to technical aspects of surface gravity.

Our conventions remain the same as in Refs. \cite{x7,x8}. The latin indices $(i,j,\dots)$ refer to the local Lorentz frame, the greek indices $(\m,\n,\dots)$ refer to the coordinate frame,  $b^i$ is the orthonormal coframe (tetrad), $h_i$ is the dual basis (frame) such that $h_i\inn b^k = \d_i^k$, the Lorentz metric is $\eta_{ij}=(1,-1,-1,-1)$, and $\om^{ij}$ is a  metric compatible (antisymmetric) connection. The exterior product of forms is implicit, the volume 4-form is $\heps = b^0b^1b^2b^3$, the Hodge dual of a form $\a$ is denoted by $\hd\a$, with $\hd 1=\heps$, and the totally antisymmetric symbol $\ve_{ijmn}$ is normalized to $\ve_{0123}=+1$.

%%%%%%%%%%%%%%%%%%%%%%%%%%%%%%%%%%%%%%%%%%%%%%%%%%%%%%%%%%%%%%%%%%%%%%%%%%%%%%%
\section{General formalism}\label{sec2}
\setcounter{equation}{0}

In the Hamiltonian approach to entropy proposed in Ref. \cite{x7}, a generalized concept of entropy is introduced in the framework of Poincar\'e gauge theory  \cite{x10,x11}. By construction, Poincar\'e gauge theory is characterized by a Riemann-Cartan geometry of the underlying spacetime continuum, in which both the torsion and the curvature determine the gravitational dynamics. Adopting the requirement of vanishing torsion, this approach can be equally well applied to Riemannian spacetimes.

To examine black hole entropy in Riemannian \grl, we start with the first-order tetrad formalism, in which basic Lagrangian variables are the coframe $b^i=b^i{_\m}d x^\m$ and the antisymmetric (metric compatible) spin connection $\om^{ij}=\om^{ij}{_\m}dx^\m$ (1-forms). These variables are a priori independent. In the absence of matter, the gravitational dynamics is determined by the Lagrangian
\be
L_G= -a_0\hd(b_ib_j)R^{ij}-2\L_0\heps\, ,
\ee
where $R^{ij}=d\om^{ij}+\om^i{_k}\om^{kj}$ is the curvature 2-form,  $a_0=1/16\pi$ (in units $G=1$), and $\L_0$ is a cosmological constant.

The field equations are derived by varying $L_G$ with respect to $b^i$ and $\om^{ij}$:
\bea
\d b^i:&&-a_0\ve_{ijmn}b^jR^{mn}-2\L_0\heps_i=0\,,                  \nn\\
\d\om^{ij}:&&-2a_0\ve_{ijmn}(\nab b^m)b^n=0\,,
\eea
where $\heps_i:=h_i\inn\heps$ and $\nab$ is the covariant derivative. The second equation implies the vanishing of torsion, $T^i:=\nab b^i\equiv db^i+\om^i{_k}b^k =0$, and consequently, $\om^{ij}$ becomes a Riemannian connection, completely determined by the coframe $b^i$. When such an $\om^{ij}$ is substituted into the first equation, the result is equivalent to Einstein's equation for \grl\ in vacuum.

The Hamiltonian approach to entropy \cite{x7} originated from the  canonical realization of the idea that entropy is the Noether charge, see Wald \cite{x4} and Jacobson et al. \cite{x5}. To clarify the idea, recall that the canonical charges are defined by certain boundary terms at spatial infinity \cite{x2}. Then, by extending the concept of charge, one can naturally define entropy by the same type of boundary terms, but located on horizon. In other words, entropy is just the \emph{canonical charge on horizon}. To be more precise, let $\S$ be a spatial section of spacetime. For spacetime with a black hole, we assume that the boundary of $\S$ consists of two components, one at infinity and the other on horizon, $\pd\S=S_\infty\cup S_H$, so that the total boundary term $\G$ has the form $\G:=\G_\infty-\G_H$ (different signs reflect opposite orientations of $S_H$ and $S_\infty$). In the case of a manifestly stationary and axially symmetric black hole in \grl, with Killing vectors $\pd_t$ and $\pd_\vphi$, the  boundary terms $\G_\infty$ and $\G_H$ are defined by the \emph{variational equations}
\bsubeq\lab{2.3}
\bea
&&\d\G_\infty=\oint_{S_\infty}\d B(\xi)\,,\qquad
       \d\G_H=\oint_{S_H} \d B(\xi)\,,                                  \\
&&\d B(\xi):=\frac{1}{2}(\xi\inn\om^{ij})\d H_{ij}
             +\frac{1}{2}\d\om^{ij}(\xi\inn H_{ij})\, ,
\eea
where $\xi$ is a Killing vector ($\pd_t$ or $\pd_\vphi$ on $S_\infty$, and a linear combination thereof on $S_H$) and $H_{ij}$ is the covariant momentum
\be
H_{ij}:=\frac{\pd L_G}{\pd R^{ij}}=-a_0\ve_{ijmn}b^mb^n\,.
\ee
\esubeq
The above form of $\d B(\xi)$ is a restriction of the general result found in \cite{x7}, to the case of Riemannian spacetimes \cite{x9}. The operation $\d$ is determined by an appropriate choice of  \emph{boundary conditions}, which ensures the variational problem to be well-defined.
\bitem
\item[(a1)] The variation $\d\G_\infty$ is realized by varying the parameters of a black hole, leaving the background configuration fixed.\vsm
\item[(a2)] The variation $\d\G_H$ is performed by varying parameters of a black hole on horizon, but keeping surface gravity constant, in accordance with the zeroth law.
\eitem
When the variational equations \eq{2.3} are $\d$-\emph{integrable} and the integrals $\G_\infty$ and $\G_H$ are \emph{finite}, they are interpreted as the asymptotic canonical charges (energy and angular momentum) and entropy, respectively.

Since $\G_\infty$ and $\G_H$ are defined as a priori independent quantities, relations \eq{2.3} do not tell us anything about the first law. However, one can show that the functional differentiability (regularity) of the canonical generator $G$, associated to the Killing vectors $\pd_t$ and $\pd_\vphi$, is ensured by going over to its improved from $\tG=G+\G$, where $\G=\G_\infty-\G_H$, see Regge and Teitelboim \cite{x2}. In other words, $\d\tG\equiv\d G+\d\G=R$, where $R$ stands for regular terms. Thus, the canonical generator $G$ is differentiable, $\d G=R$, if
\be
\d\G\equiv\d\G_\infty-\d\G_H=0\,,                                  \lab{2.4}
\ee
which is nothing but the \emph{first law} of black hole dynamics in its variational form \cite{x12}.

%%%%%%%%%%%%%%%%%%%%%%%%%%%%%%%%%%%%%%%%%%%%%%%%%%%%%%%%%%%%%%%%%%%%%%%%%%%%%%%
\section{Tetrad form of Kerr-AdS geometry}\label{sec3}
\setcounter{equation}{0}

Kerr-AdS metric in Boyer-Linquist coordinates can be written in the form \cite{x13,x6}
\bsubeq\lab{3.1}
\be
ds^2=\frac{\D}{\r^2}\Big(dt+\frac{a}{\a}\sin^2\th d\vphi\Big)^2-\frac{\r^2}{\D}dr^2
  -\frac{\r^2}{f}d\th^2
 -\frac{f}{\r^2}\sin^2\th\Big[a dt+\frac{r^2+a^2}{\a}d\vphi\Big]^2\,,
\ee
where
\bea
&&\D:=(r^2+a^2)(1+\l r^2)-2mr\, ,\qquad \a:=1-\l a^2\,,               \nn\\
&&\r^2:=r^2+a^2\cos^2\th\,,\qquad f:=1-\l a^2\cos^2\th\,.
\eea
\esubeq
The metric is an exact solution of \grl\ for $3a_0\l+\L_0=0$, $m$ and $a$ are parameters of the solution, and $\a$ is introduced to normalize the range of $\vphi$ to $2\pi$. The metric has two Killing vectors, $\pd_t$ and $\pd_\vphi$, and the outher horizon is located at the larger root of $\D(r)=0$:
\be
(r_+^2+a^2)(1+\l r_+^2)-2mr_+=0\,.
\ee
The line element \eq{3.1} defines the metric components
\bsubeq
\bea
&&g_{tt}=\frac{1}{\r^2}(\D-fa^2\sin^2\th)\,,\qquad
  g_{t\vphi}=-\frac{a}{\r^2\a}\sin^2\th\Big[f(r^2+a^2)-\D\Big]\, ,      \nn\\
&& g_{\vphi\vphi}=-\frac{\S^2}{\r^2\a^2}\sin^2\th \,,
\eea
where
\be
\S^2:=f(r^2+a^2)^2-a^2\D\sin^2\th\, ,
\ee
\esubeq
The angular velocity is given by
\be
\om:=\frac{g_{t\vphi}}{g_{\vphi\vphi}}
    =\frac{a\a\big[f(r^2+a^2)-\D\big]}{f(r^2+a^2)^2-a^2\D\sin^2\th}\, ,
\qquad\om_+:=\om(r_+)=\frac{a\a}{r_+^2+a^2}\, .                    \lab{3.4}
\ee
For large $r$, the angular velocity does not vanish, $\om=-\l a+O_2$.
Surface gravity is given by (Appendix \ref{appA}):
\be
\k=\frac{(\pd_r\D)_{r_+}}{2(r_+^2+a^2)}=
   \frac{r_+\big(1+\l a^2+3\l r_+^2-a^2/r_+^2\big)}{2(r_+^2+a^2)}\,.
\ee

The form of the metric \eq{3.1} suggests the following orthonormal coframe:
\bsubeq\lab{3.6}
\bea
&&b^0=N\Big(dt+\frac{a}{\a}\sin^2\th\,d\vphi\Big)\,,\qquad
  b^1=\frac{dr}{N}\,,\nn                                               \\
&&b^2=Pd\th\, ,\qquad
b^3=\frac{\sin\th}{P}\Big[a\,dt+\frac{r^2+a^2}{\a}d\vphi\Big]\,,
\eea
where
\be
N=\sqrt{\D/\r^2}\, ,\qquad P=\sqrt{\r^2/f}\, .
\ee
\esubeq
The horizon area is given by
\be
A=\int_{r_+} b^2b^3=4\pi\frac{r_+^2+a^2}{\a}\, .
\ee
The Riemannian connection has the form
\bea
&&\om^{01}=-N'b^0-\frac{ar}{P\r^2}\sin\th b^3\,,\qquad
\om^{02}
  =\frac{a^2\sin\th\cos\th}{P\r^2}b^0-\frac{aN}{\r^2}\cos\th b^3\,,    \nn\\
&&\om^{03}=-\frac{ar}{P\r^2}\sin\th b^1+\frac{aN}{\r^2}\cos\th b^2\,, \nn\\
&&\om^{12}=\frac{a^2\sin\th\cos\th}{P\r^2}b^1+\frac{Nr}{\r^2}b^2\,,\qquad
  \om^{13}=-\frac{ar}{P\r^2}\sin\th b^0+\frac{Nr}{\r^2}b^3\,,         \nn\\
&&\om^{23}=-\frac{aN}{\r^2}\cos\th b^0
            +\frac{P\cos\th-\pd_\th P\sin\th}{P^2\sin\th}b^3\, .   \lab{3.8}
\eea
There are two nonvanishing irreducible parts of the curvature 2-form $R^{ij}=d\om^{ij}+\om^i{}_k\om^{kj}$, $\ir{1}R^{ij}$ and $\ir{6}R^{ij}$, see for instance \cite{x7,x11}. The Weyl curvature $W^{ij}\equiv\ir{1}R^{ij}$ is given by \cite{x14}
\bsubeq\lab{3.9}
\bea
&&W_{01}=2Cb^0b^1+2D b^2b^3\,,\qquad W_{12}= Cb^1b^2-Db^0b^3 \,,\nn\\
&&W_{02}=-Cb^0b^2+Db^1b^3\, ,\qquad W_{13}= Cb^1b^3 +D b^0b^2 \,, \nn\\
&&W_{03}=-Cb^0b^3 -D b^1b^2\,,\qquad W_{23}=-2Cb^2b^3+2D b^0b^1\, .
\eea
where the coefficients $C$ and $D$ are the same as in Kerr spacetime  \cite{x13,x15}:
\be
C:=\frac{mr}{\r^6}(r^2-3a^2\cos^2\th)\, ,\qquad
D:=\frac{ma\cos\th}{\r^6}(3r^2-a^2\cos^2\th)\, .
\ee
\esubeq
The irreducible part $\ir{6}R^{ij}$, Ricci curvature and the curvature scalar read
\be
\ir{6}R^{ij}:=R^{ij}-W^{ij}=\l b^ib^j\,,\qquad \ric^i=3\l b^i\, ,\qquad R=12\l\,.
\ee
The quadratic curvature invariant,
\be
R^{ij}\hd R_{ij}=\frac{24 m^2}{\r^{12}}\big(r^2-a^2\cos^2\th\big)
                          \big(\r^4-16a^2r^2\cos^2\th\big)\heps\,,
\ee
exhibits a singularity at $\r^2=0$, but not at the horizon $r=r_\pm$. For $m=0$, the spacetime is described by the AdS geometry in \emph{non-standard coordinates}, in which metric components depend on the parameter $a$.

When the above geometric structure is combined with the variational equations \eq{2.3}, it allows us to find asymptotic charges and entropy of Kerr-AdS black holes, as well as to verify the validity of the first law \eq{2.4}.

%%%%%%%%%%%%%%%%%%%%%%%%%%%%%%%%%%%%%%%%%%%%%%%%%%%%%%%%%%%%%%%%%%%%%%%%%%%%%%%
\section{Asymptotic charges}\label{sec4}
\setcounter{equation}{0}

\subsection{Inadequacy of Boyer-Lindquist coordinates}

As noted by Henneaux and Teitelboim \cite{x3}, see also Carter \cite{x13}, one cannot properly define asymptotic conditions for Kerr-AdS metric in Boyer-Lindquist (BL) coordinates. Although our approach relies on a different, variational procedure, it also leads, as we shall see, to certain problems with BL coordinates, which are essentially caused by the non-standard description of the background geometry. Nevertheless, it is instructive to see how our approach works in the BL coordinates for two reasons: first, it will help us to identify type of the related problems, and second, it will indicate a way towards a proper solution.

To begin with calculations, we need to precisely define the variational procedure. The background geometry is defined by $m=0$, but it continues to depend on the parameter $a$ that appears also in the black hole solution. On the other hand, the rule (a1) from section \ref{sec2} requires to avoid varying over those $a^\prime$s that are associated to the background. How can one practically realize this rule? The answer is given by the following simple rule \cite{x9}:
\bitem
\item[(a3)] In Eq. \eq{2.3}, first apply $\d$ to all the parameters $(m,a)$ appearing in $B(\xi)$, then subtract the terms that survive the limit $m=0$, as they stem from the variation of the AdS background. Thus, symbolically, $\d B(\xi):=\d_{(m,a)}B(\xi)-\d_{(0,a)} B(\xi)$.
\eitem
Applying this rule to the calculation of energy, one finds the following nonvanishing contributions to $\d\G_\infty[\pd_t]$ (integration is postponed, $m$-independent terms are subtracted):
\bea
\om^{01}{}_t\d H_{01}&=&
   -2a_0\big(\frac{m\d\a}{\a^2}\Big)d\Om\, ,                         \nn\\
\d\om^{12}H_{12t}
   &=&2a_0\frac{\d m f- m\d f}{\a f}d\Om\,,                          \nn\\
\d\om^{13}H_{13t}
   &=&2a_0\frac{\a(\d m f +m\d f)-2\d\a fm}{\a^2 f}d\Om\,,
\eea
where  $d\Om:=\sin\th d\th d\vphi$. Summing up and integrating over the spatial boundary $S_\infty$ yields
\be
\d E_t:=\d\G_\infty[\pd_t]
  =\frac{m}{2}\d\Big(\frac{1}{\a}\Big)+\d\Big(\frac{m}{\a}\Big)\,, \lab{4.2}
\ee
where we used $16\pi a_0=1$. We face here a serious problem, intrinsically related to the BL coordinate system:
\bitem
\item[S1.] The variational equation \eq{4.2} is not $\d$-integrable, and without that, the calculation of energy doesn't make sence.
\eitem

An analogous calculation of $\d\G_\infty[\pd_\vphi]$ leads to
\bea
&&\d\om^{01}H_{01\vphi}+\om^{01}{}_\vphi\d H_{01}
  =2a_0\d\Big(\frac{ma}{\a^2}\Big)d\Om'\, ,                           \nn\\
&&\d\om^{13}H_{13\vphi}+\om^{13}{}_\vphi\d H_{13}
  =4a_0\d\Big(\frac{ma}{\a^2}\Big)d\Om'\, ,
\eea
where $d\Om':=\sin^3\th d\th d\vphi$. The resulting expression for the angular momentum is integrable:
\be
\d E_\vphi:=\d\G_\infty[\pd_\vphi]=\d\Big(\frac{ma}{\a^2}\Big)\,.   \lab{4.3}
\ee

\subsection{Untwisting the BL coordinate system}

In their analysis of asymptotically Kerr-AdS spacetimes, Henneaux and Teitelboim \cite{x3} introduced boundary conditions in terms of the asymptotic metric configurations which satisfy three main requirements: the set of asymptotic states should contain Kerr-AdS solution, it must be invariant under the $SO(2,3)$ group of asymptotic symmetries, and the associated boundary terms should be well-defined and finite. Since the metric in the BL coordinates does not obey these requirements, the authors used an implicit coordinate transformation
\bsubeq\lab{4.5}
\bea
&&T=t\,,\qquad \phi=\vphi-\l at\,,                               \lab{4.5a}\\
&&R\cos\Th=r\cos\th\, ,\qquad \a R^2=r^2f+a^2\sin^2\th\,,
\eea
\esubeq
which untwists the metric \eq{3.1} to the standard asymptotically AdS form; see also Carter \cite{x13}. The asymptotic states are not explicitly used in our variational approach, as their role is effectively taken over by the rule (a3). However, the problem with $\d$-non-integrability of $\d E_t$  is a signal to reconsider the use of the BL coordinates. It turns out that the problem can be resolved by using only the $(T,\phi)$ part \eq{4.5a} of the coordinate transformations \eq{4.5}. Indeed, as we shall see, the $(T,\phi)$ transformations are \emph{sufficient} to untwist the BL frame in a way which ensures $\d$-integrability of both asymptotic charges.

The vector and tensor components in the $(T,\phi)$ frame are obtained from their BL expressions by the standard law of coordinate transformations:
\bea
&&\xi_T=\xi_t+\l a\xi_\vphi\, ,\qquad \xi_\phi=\xi_\vphi\, ,        \nn\\
&&g_{T\phi}=g_{t\vphi}+\l a g_{\vphi\vphi}\, ,\qquad
  g_{\phi\phi}=g_{\vphi\vphi}\,,                                    \nn\\
&&g_{TT}=g_{tt}+2\l a g_{t\vphi}+(\l a)^2g_{\vphi\vphi}\,.       \lab{4.6}
\eea
As a consequence, the new angular velocity is found to be
\be
\Om=\frac{g_{T\phi}}{g_{\phi\phi}}=\om+\l a\, ,\qquad
\Om_+=\om_++\l a=\frac{a(1+\l r_+^2)}{r_+^2+a^2}\,.
\ee
In contrast to the old expression \eq{3.4}, $\Om$ vanishes at infinity, which is why the $(T,\phi)$ frame is often referred to as non-rotating. In a similar manner, the new components of the coframe and connection 1-forms are obtained from the BL expressions \eq{3.6} and \eq{3.8}:
\bea
&&b^i{_T}=b^i{_t}+\l a b^i{_\vphi}\,,\qquad b^i{_\phi}=b^i{_\vphi}\,,\nn\\
&&\om^{ij}{_T}=\om^{ij}{_t}+\l a\om^{ij}{_\vphi}\,,\qquad
  \om^{ij}{_\phi}=\om^{ij}{_\vphi}\,.                            \lab{4.8}
\eea

Now, we are ready to calculate the asymptotic charges. Starting from the variational equations \eq{2.3}, energy and angular momentum are defined by the relations
\be
\d E_T:=\d\G_\infty[\pd_T]\, ,\qquad \d E_\phi:=\d\G_\infty[\pd_\phi]\,,
\ee
respectively. These definitions, combined with the relations \eq{4.8}, make the final step surprisingly simple:
\bsubeq
\bea
&&\d E_T=\frac{1}{2}\om^{ij}{}_T\d H_{ij}+\frac{1}{2}\d\om^{ij}H_{ijT}
      =\d E_t+\l a\d E_\vphi=\d\Big(\frac{m}{\a^2}\Big)\, ,     \lab{4.10a}\\
&&\d E_\phi=\frac{1}{2}\om^{ij}{}_\phi\d H_{ij}+\frac{1}{2}\d\om^{ij}H_{ij\phi}
      =\d E_\vphi=\d\Big(\frac{m}{\a^2}\Big)\, .
\eea
\esubeq
Since both relations are $\d$-integrable, the corresponding asymptotic charges are
\be
E_T=\frac{m}{\a^2}\, ,\qquad E_\phi=\frac{ma}{\a^2}\,.            \lab{4.11}
\ee

\bitem
\item[S2.] The asymptotic charges are well defined in the $(T,\phi)$ frame which is non-rotating at infinity with respect to the AdS background.
\eitem
Thus, the problem with $\d$-non-integrability of $\d E_t$ is solved by going over to the $(T,\phi)$ frame. The charges \eq{4.11} coincide with those obtained originally by Henneaux and Teitelboim \cite{x3}, and confirmed in a different context by Hecht and Nester \cite{x16}; see also Refs. \cite{x17,x18}.

%%%%%%%%%%%%%%%%%%%%%%%%%%%%%%%%%%%%%%%%%%%%%%%%%%%%%%%%%%%%%%%%%%%%%%%%%%%%%%%
\section{Entropy and the first law}\label{sec5}
\setcounter{equation}{0}

In the $(T,\phi)$ coordinate system, entropy is determined by the variational equation for $\G_H(\xi)$ with $\xi=\pd_T-\Om_+\pd_\phi$. Starting with the relations
\bsubeq
\bea
&&N\pd_r N\big|_{r=r_+}=\frac{\k(r_+^2+a^2)}{\r_+^2}\, ,                \nn\\
&&\xi\inn b^0\big|_{r_+}=\frac N\a(f-\Om_+a\sin^2\th)\, ,\qquad
  \xi\inn b^a\big|_{r_+}=0\,,
\eea
where $a=1,2,3$, one finds
\bea
&&\xi\inn H_{ij}\big|_{r=r_+}=0\, ,                             \nn\\[3pt]
&&\xi\inn\om^{01}\big|_{r=r_+}
                    =-\frac{N\pd_r N\big|_{r=r_+}}\a(f-\Om_+a\sin^2\th)=-\k\,,
\eea
\esubeq
whereas the remaining $\xi\inn\om^{ij}$ terms vanish. Hence, there is only one nonvanishing term contributing to the Kerr-AdS entropy:
\bea
&&(\xi\inn\om^{01})\d H_{01}=2a_0\k
  \d\Big(\frac{r_+^2+a^2}\a\Big)\sin\th d\th d\phi\,,               \nn\\
&&\d\G_H[\xi]=\oint_H(\xi\inn\om^{01})\d H_{01}
             =8\pi a_0\k\d\Big(\frac{r_+^2+a^2}\a\Big)\, ,
\eea
so that
\be
\d\G_H[\xi]=T\d S\,,\qquad S:=\pi\frac{r_+^2+a^2}\a\, ,
\ee
where  $T:=\k/2\pi$ is the black hole temperature.

A direct comparison of $\d\G_H$ to $\d\G_\infty$ yields $\d\G_\infty[\xi]=\d\G_H[\xi]$, in accordance with the general relation \eq{2.4}. In other words:
\bitem
\item[S3.] The thermodynamic variables $(E_T,E_\phi,S)$ satisfy the first law of black hole dynamics:
\be
\d E_T-\Om_+\d E_\phi=T\d S\,.
\ee
\eitem

%%%%%%%%%%%%%%%%%%%%%%%%%%%%%%%%%%%%%%%%%%%%%%%%%%%%%%%%%%%%%%%%%%%%%%%%%%%%%%%
\section{Concluding remarks}\label{sec6}
\setcounter{equation}{0}

In the present paper, we studied entropy and asymptotic charges for Kerr-AdS black holes in \grl, using the general Hamiltonian approach proposed in Ref. \cite{x7}. The thermodynamic variables are first consistently defined via the variational equations \eq{2.3} in the $(T,\phi)$ frame, and then, they are shown to satisfy the first law of black hole dynamics.

Controversial aspects of the first law of black hole dynamics for Kerr-AdS
solutions have been discussed in detail by Gibbons et al. \cite{x6}. To compare our results with those existing in the literature, we use the Euclidean action formalism to express the value of the action integral in the form $I_4=T^{-1}F$, where the free energy $F$ is given by \cite{x6,x19,y20}
\be
F=\frac{1}{2\a}\Big[m-\l r_+(r_+^2+a^2)\Big]
    =\frac{(r_+^2+a^2)(1-\l r_+^2)}{4\a r_+}\, .
\ee
One can directly verify that our thermodynamic variables $V=(E_T,E_\phi,\Om_+,S)$  satisfy the quantum statistical relation
\be
F=E_T-\Om_+E_\phi-TS\, ,                                          \lab{6.3}
\ee
whereas the Smarr formula \cite{x21} fails to hold for $\l\ne 0$:
\be
\frac{1}{2}E_T-\Om_+E_\phi-TS
   =-\frac{\l(r_+^2+a^2)(r_+^2+\a r_+^2+a^2)}{4\a^2 r_+}\, .      \lab{5.4}
\ee
However, treating the cosmological constant as a new thermodynamic variable analogous to pressure, one can formulate a generalized Smarr formula,
see Altamirano et al. \cite{x22}.

Hawking et al. \cite{x23} use a different set of thermodynamic variables:
\be
V'=(E',E_\phi,\om_+,S)\,,\qquad E':=\frac{m}{\a}\, ,
\ee
where $E'$ is different from $E_T$, and $\om_+$ refers to the BL frame. It is interesting that these variables also satisfy the quantum statistical relation
\be
F=E'-\om_+E_\vphi-TS\, .                                        \lab{6.5}
\ee
However, the validity of the quantum statistical relation does not guarantee the consistency of the variables that satisfy it. Namely, although the expression $E'$ can be \emph{formally} calculated in the BL frame, see Refs. \cite{x23,x24,x25}, the result does not have an acceptable physical interpretation, since energy is not a well-defined concept in the BL frame; see section \ref{sec4}, or the arguments given in Ref. \cite{x3}. It is therefore not surprising that the variables $V'$ \emph{do not satisfy} the first law \cite{x6}.

We expect that the Hamiltonian approach used in the present paper can be consistetly extended to Kerr-AdS black holes in Poincar\'e gauge theory \cite{x26,x27}.

%%%%%%%%%%%%%%%%%%%%%%%%%%%%%%%%%%%%%%%%%%%%%%%%%%%%%%%%%%%%%%%%%%%%%%%%%%%%%%%
\section*{Acknowledgments}

This work was partially supported by the Serbian Science Foundation under Grant No. 171031.

\appendix
%%%%%%%%%%%%%%%%%%%%%%%%%%%%%%%%%%%%%%%%%%%%%%%%%%%%%%%%%%%%%%%%%%%%%%%%%%%%%%%
\section{Surface gravity}\label{appA}
\setcounter{equation}{0}

\def\hN{{\hat N}}

To calculate surface gravity of the Kerr-AdS black hole, it is convenient to represent its metric in the \emph{generic} form, based on the ADM variables $\hN$ and $\hN_\vphi$:
\bsubeq\lab{A.1}
\bea
ds^2&=&\hN^2dt^2-\frac{dr^2}{F^2}+g_{\vphi\vphi}(d\vphi+\hN_\vphi dt)^2
                                          +g_{\th\th}d\th^2\,,        \nn\\
  &=&g_{tt}dt^2-\frac{dr^2}{F^2}+2g_{t\vphi}dtd\vphi
                     +g_{\vphi\vphi}d\vphi^2+g_{\th\th}d\th^2\,,
\eea
where $x^\m=(t,r,\th,\vphi)$ are Schwarzschild-like coordinates, and
\be
\hN_\vphi=\frac{g_{t\vphi}}{g_{\vphi\vphi}}\, , \qquad \hN^2=g_{tt}-\frac{g^2_{t\vphi}}{g_{\vphi\vphi}}\, .
\ee
\esubeq
We assume that $F$ vanishes on horizon, $F(r_+)=0$, but $g:=\hN/F$  does not, $g(r_+)\ne 0$. Performing the coordinate transformation
\be
dv=dt+\frac{1}{gF^2}\,dr\, ,\qquad d\psi=d\vphi-\frac{\hN_\vphi}{gF^2}\,dr\,,                         \ee
one can bring the metric \eq{A.1} to the Edington-Finkelstein form
\be
ds^2=g_{tt}dv^2-2gdvdr+2g_{t\vphi}dvd\psi+g_{\vphi\vphi}d\psi^2
                                         +g_{\th\th}d\th^2\,,
\ee
which is regular on horizon. Then, using the Killing vector $\xi=\pd_v-\om_+\pd_\psi$, surface gravity can be calculated from the relation $\pd_\m\xi^2=-2\k\xi_\m$ at horizon:
\bea
&&\xi^2=g_{\vphi\vphi}(\om-\om_+)^2+\hN^2\,,\qquad
                                    \xi_\m=(0,-g,0,0)\, ,   \nn\\[3pt]
&&\Ra\qquad \k=F\pd_r\hN\big|_{r_+}\,.
\eea

Applying this general result in the BL frame, one obtains
\be
F^2=\frac{\D}{\r^2}\, ,\quad
\hN^2=\frac{f\r^2}{\S^2}\D\quad\Ra\quad
   \k=\frac{(\pd_r\D)_{r_+}}{2(r_+^2+a^2)}\, .
\ee
On the other hand, the identity
\be
g_{tt}-\frac{g_{t\vphi}^2}{g_{\vphi\vphi}}
                          =g_{TT}-\frac{g_{T\phi}^2}{g_{\phi\phi}}
\ee
ensures that the value of $\hN^2$, as well as of $\k$, remains the same in the $(T,\phi)$ frame.

%%%%%%%%%%%%%%%%%%%%%%%%%%%%%%%%%%%%%%%%%%%%%%%%%%%%%%%%%%%%%%%%%%%%%%%%%%%%%%%


\begin{thebibliography}{99}
\bibitem{x12}  The inverse statement starts from $\d G=R$. Then, since $G$ is differentiable, one can choose $\d\G=0$ without loss of generality, and the inverse statement is essentially trivial.
\end{thebibliography}
\end{document}